\documentclass[preprint,showpacs,preprintnumbers,amsmath,amssymb]{revtex4}
\usepackage{graphicx}% Include figure files
\usepackage{floatflt}
\usepackage{dcolumn}% Align table columns on decimal point
\usepackage{bm}% bold math
\usepackage{rotating}
\usepackage{natbib}
\usepackage{txfonts}
\begin{document}
 \title{ Energy systematics of heavy nuclei -- mean field models in comparison}

   \author{P. -G. Reinhard$^1$}
   \author{B. K. Agrawal$^2$}
              \email{bijay.agrawal@saha.ac.in}
   \affiliation{
   $^1$Institut f¨ur Theoretische Physik II, Universit¨at
Erlangen-N¨urnberg, Staudtstrasse 7, D-91058 Erlangen, Germany\\
$^2$Saha Institute of Nuclear Physics, Kolkata - 700064,
India.}
\date{\today}

\preprint{APS/123-QED}

\begin{abstract}
We compare the systematics of binding energies computed within the
standard and extended versions of the relativistic mean-field (RMF)
model and the Skyrme Hartree-Fock (SHF)  model. The general trends for
the binding energies for super-heavy nuclei are significantly different
for these models.  The SHF models tend to underbind the superheavy nuclei,
while, RMF models show just  the opposite trend.  The extended RMF model
seems to provide remarkable improvements over the results obtained for the
standard RMF model.  
 \end{abstract}  \maketitle
\section{Introduction}
Self-consistent mean field (SCMF) models are the most feasible means
for the microscopic description of ground state properties and
low-energy collective dynamics of nuclei. These models are also
employed for the study of compact stars, since they can be easily
extended to include the contributions from hyperons and exotic
phenomena as, e.g., kaon condensation \cite{Gle00aB,Sto07aR}. The
three most prominent SCMF models are the Skyrme Hartree-Fock approach
(SHF) \cite{Vau72a,Ben03aR}, the Gogny force \cite{Dec80a}, and the
relativistic mean field model (RMF) \cite{Ser86aR,Rei89aR,Rin96aR}.
We shall focus on SHF and RMF in comparison. SHF, as the name
suggest, is based on the Skyrme energy functional derived from a
zero-range effective interaction.  The RMF models are
based on an effective Lagrangian density which describes the
interactions of nucleons through the exchange of the scalar-isoscalar
($\sigma$), vector-isoscalar ($\omega$) and vector-isovector ($\rho$)
meson fields.  

The form of the energy functional or Lagrangian density is given by
invariance constraints and plausible physical arguments (as, e.g.,
low-momentum expansion \cite{Vau72a,Rei94aR,Ben03aR}). However, the
parameters of the models are free parameters. A strict derivation from
a microscopic nucleon-nucleon interaction has not yet been too
successful at a quantitative level because ab-initio models, in spite
of the enormous success over the last years, have still a limited
quality if one refrains from adding empirical corrections like a
three-body force (see, e.g., \cite{Nav00a,Dea05a,Nav09aR}). One of the
important tasks in the development of mean field models is thus to
devise ways of constraining the free parameters of interaction
used. This is done usually by an adjustment to empirical data. This
is, however, rather involved as one does not have a one-to-one
correspondence between individual parameters and a specific piece of
experimental information. Some parameters are well fixed by known
nuclear ground state properties. Others remain only loosely determined
leaving leeway for additional data or bias. Therefore, many tens of
different parameter sets for SHF and RMF have evolved in the course of
the development.  A key task is to figure out which features depend on
a particular parameterization and which features are more related to
the intrinsic structure of the model.  Recent extensive explorations
of the SHF functional have indicated a few of such intrinsic features
\cite{Klu09a,Erl10a}. One key problem is that all SHF functionals show
a systematic trend to underbinding for super-heavy elements while
traditional RMF parameterizations show an opposite trend to
overbinding. It is the aim of the present contribution to study this
discrepancy in more detail, in particular with more data and for a
broader selection at the side of the RMF.  To this end, we shall
compare the binding energy systematics for even-even nuclei computed
using the standard and extended versions of the RMF models and the SHF
models.

The RMF can be grouped into three different classes: (1) models with
non-linear meson self-coupling \cite{Ser86aR}, (2) models with
density-dependent nucleon-meson coupling \cite{Typ99a,Vre05a}, (3)
models which combine (1) and (2).  We concentrate here on class (1).
Within these non-linear RMF one can distinguish two sub-classes: (i)
standard RMF which includes the contributions from the the non-linear
self-couplings for the $\sigma$ meson field only, and (ii) extended
RMF model which includes the contributions from the self and/or cross
interaction terms for the $\sigma$, $\omega$ and $\rho$ meson
fields. We will discuss both versions of meson coupling.
\section{Choices for the models}

We have considered four different forces for both the RMF and
SHF models.  The specific reasons for choosing these forces are as
follows.  In particular, we employ the RMF parameterizations:
NL3 \cite{Lal97a}, TM1 \cite{Sug94a}, FSUGold \cite{Tod05a}, and
BSR4 \cite{Dhi07a, Agr10a}.  The non-linear part of the Lagrangian
density  for these RMF models are not the same. 
These differences are summarized in Table
\ref{tab:rmf}. The Lagrangian density associated with the parameter
set NL3 which represents the standard RMF model contains non-linear
term only for the $\sigma$ meson self-interaction.  The Lagrangian
density corresponding to the parameter set TM1 is extended to include
the self-interaction term also for the $\omega$ mesons.  The FSUGold
parameter set further includes the $\omega-\rho$ cross-interaction
term. The Lagrangian density corresponding to the BSR4 parameter
set includes all the cross-interaction terms for $\sigma$, $\omega$
and $\rho$ mesons up to quartic order, but, the self-interaction of the
$\omega$-meson is switched off. These parameter sets should enable us
to delineate the effects of the meson self- and cross-interactions on
the binding energy systematics.  We neglect the contributions from the
self interaction of $\rho$ mesons as they can affect the properties of
the finite nuclei and neutron stars only marginally \cite{Mue96a}.

At the side of SHF, we consider the following four parameterizations: SkI3
\cite{Rei95a}, SLy6 \cite{Cha98a} , BSk4 \cite{Gor03a} and SV-min
\cite{Klu09a}. The form of the SHF functional is basically the same
for all four cases with the tiny difference that SkI3 and SV-min
include an isovector contribution to the spin-orbit term
\cite{Rei95a}.  The parameterizations differ mainly in the bias and
choice of data for the phenomenological adjustment. The two older
parameterizations SkI3 and SLy6 used a rather small set of fit
nuclei. SkI3 included the electro-magnetic form factor and isotope
shifts as data while SLy6 put emphasis on the neutron rich side and
included data from bulk matter. BSk4 puts emphasis almost exclusively
on binding energies, and includes all known even-even nuclei in the
data base. SV-min takes a data set of good ``mean-field nuclei'' which
were carefully selected to contain only negligible corrections from
collective correlations \cite{Klu08a}. This amounted to long chains of
semi-magic nuclei.  The fit set is much larger than the one for SkI3
and SLy6, but much smaller than for BSk4. On the other hand, SV-min
takes care not only of binding energies but of a larger set of
observables (radii, electro-magnetic form factor, spin-orbit
splittings, ...). The four selected forces are thus sufficiently
different with respect to these aspects of bias and choice.

We would like to briefly outline  the manner in which the
corrections to the binding energies arising from  the centre of mass motion,
pairing and quadrupole correlations are incorporated.
All models include the correction for the spurious center-of-mass energy
$E_\mathrm{cm}$. The SHF sets compute it from,
\begin{equation}
E_\mathrm{cm}=\langle\hat{P}_\mathrm{cm}^2\rangle/(2mA),
\label{eq:ecm1}
\end{equation}
while the RMF sets use the estimate 
\begin{equation}
E_\mathrm{cm}=31\,A^{-1/3}\,\mathrm{MeV}.
\label{eq:ecm2}
 \end{equation}
The values of $E_{\rm cm}$ computed using Eqs. (\ref{eq:ecm1} and
\ref{eq:ecm2}) for the cases with
mass number $A \geqslant 200$, which is of present interest,  
may differ at most by $\sim 0.5$ MeV \cite{Bender00}.  The pairing
is treated within the SHF models  using a (density dependent) zero-range
force and smooth cutoff in pairing space \cite{Bon85a,Kri90a,Ben00c}.
For the  RMF models, the contributions from the pairing correlations
to the binding energy are evaluated in the constant gap approximation
with the gap \cite{Rin80aB}, 
 \begin{equation} \Delta=\frac{11.2}{\sqrt A}\,\mathrm{MeV.}
  \label{eq:pair} \end{equation}
Soft nuclei and deformed nuclei can develop substantial contributions
from quadrupole correlations \cite{Klu08a}.  As an efficient and simple
estimate we include a large part of these correlations in terms of
approximate rotational projection expressed through variance of total
angular momentum $\langle\hat{J}^2\rangle$ and moment of inertia
$\Theta$ \cite{Hag03a,Klu08a}. This is done so far for the SHF models. In
case of RMF, we use a simple estimate for the quadrupole correlation
energy or rotational correction as, 
\begin{equation} \label{eq:erot}
 E_\mathrm{rot}=2.2\sqrt{\beta_2-0.05}\,m/m^*\,\mathrm{MeV} 
  \end{equation}
where $m^*$ is the nucleon effective mass in bulk equilibrium matter.
This estimate has been extracted by studying the microscopically computed
trends of $E_\mathrm{rot}$ for a wide variety of SHF parameterizations.
Thus, the manner in which the contributions from the c.m. correction
$E_\mathrm{cm}$, from the pairing correlations and from the rotational
correction $E_{\rm rot}$ to the binding energy evaluated for the SHF and
the RMF models are not exactly the same.  We have checked these variants
extensively and found that they do not affect the general trends which
we aim to discuss.  To this end, we may point out that the contributions
from the  Fock and the Coulomb exchange terms are ignored in the RMF
models, but, included in the SHF models.

\section{Binding energy systematics}
\begin{figure*}
\resizebox{6.1in}{!}{ \includegraphics[]{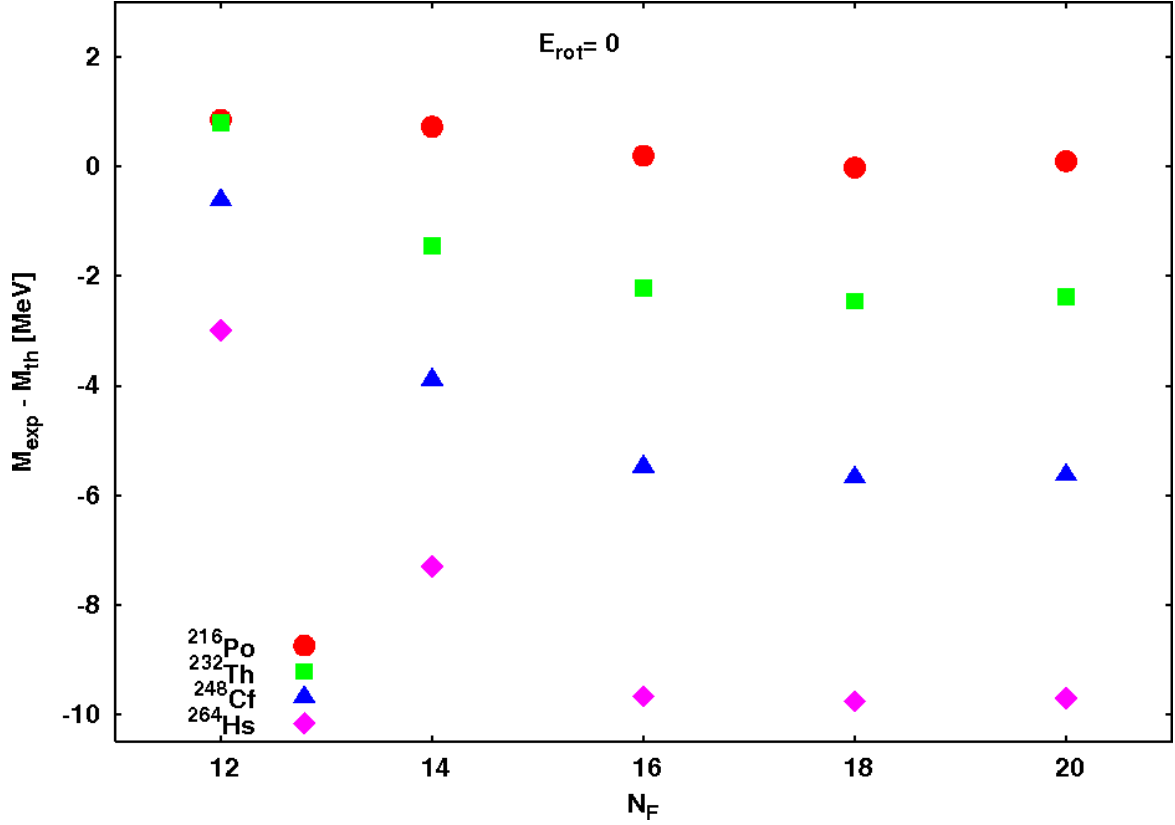}}
\caption{\label{fig:zpe0} (color online)
Errors in the binding energy (Eq. \ref{eq:delM}) plotted as a function
of the number of oscillator shells $N_F$ employed to expand the Dirac
spinors for nucleons.
 }
\end{figure*}

\begin{figure*}
\resizebox{6.1in}{!}{ \includegraphics[]{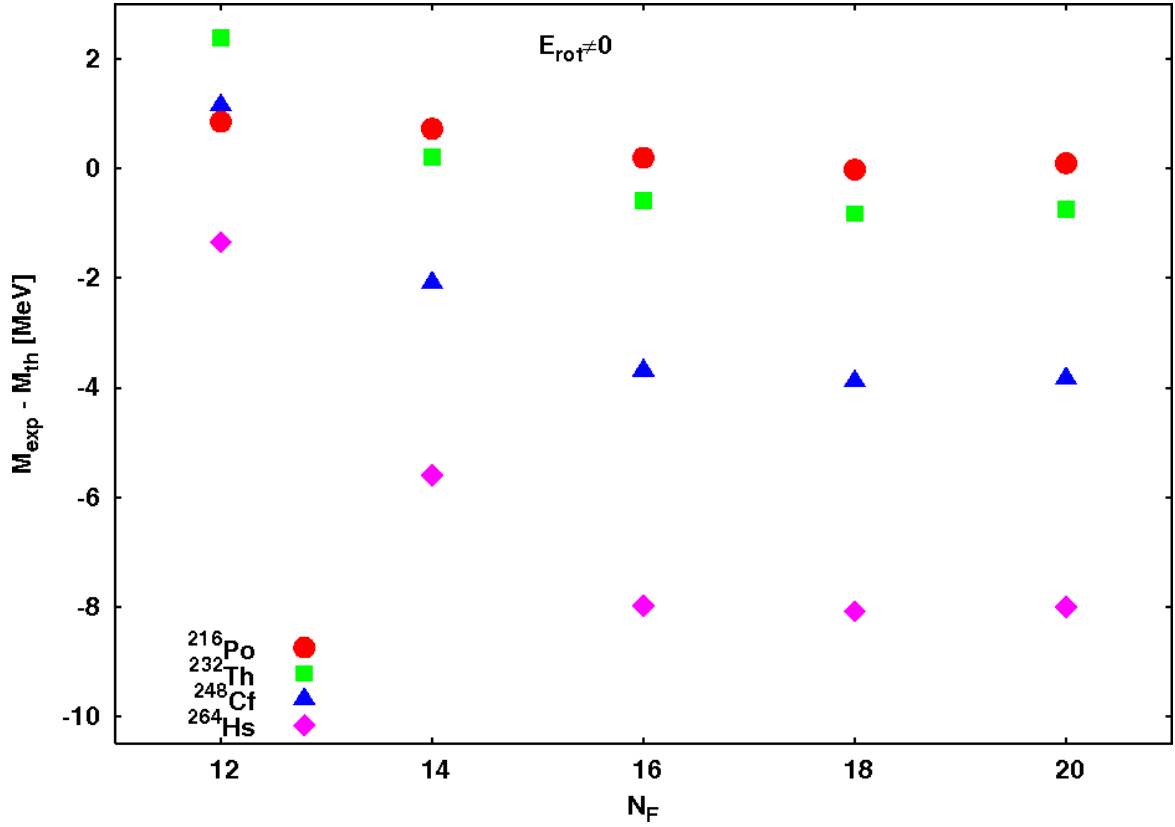}}
\caption{\label{fig:zpe} (color online)
Same as Fig. \ref{fig:zpe0}, but, the quadrupole correlation corrections
also included (Eq. \ref{eq:erot}).
 }
\end{figure*}

We first present some numerical details before embarking on our main
results. The RMF equations for nucleons are solved by expanding the Dirac
spinors in terms of the wave functions for the axially deformed harmonic
oscillator potential \cite{Gambhir90,Ring97}.  Similar strategy is
employed for solving the field equations for the mesons. The basis space
for the nucleons and the mesons are truncated at the major oscillator
shells $N_F$ and $N_B$, respectively.  The reliable solutions for the
field equations can be obtained with $N_B = 10$ \cite{Gambhir90}. However,
the appropriate choice of $N_F$ depends on the mass number of the
nucleus. For instance, reliable binding energy for $^{208}$Pb requires
number of oscillator shells $N_F \geqslant 12$.  It is quite natural
to expect the need for larger basis space for the case of superheavy
nuclei which are also well deformed in general.  Our prime interest is
to look into the binding energy systematics of the superheavy nuclei,
it is therefore necessary to ensure that the basis space used is sufficiently
large enough to yield meaningful  values for the binding energies for
these nuclei.  In Figs. \ref{fig:zpe0}  and \ref{fig:zpe} we display the
results for the binding energy error or equivalently the mass difference
between the measured and the calculated ones,
 \begin{equation}
\delta M = M_{\rm exp} - M_{\rm th} \label{eq:delM}
 \end{equation}
as a function of $N_F$ for some heavy nuclei. The values of $E_{\rm
rot}$ as used for the results plotted in Fig. \ref{fig:zpe} are
obtained from Eq.  (\ref{eq:erot}). It can be seen from these figures
that we need at least $N_F = 18$ to generate reliable binding energy
systematics over wide range of mass number.  In Table \ref{tab:comp}
we compare our results for the binding energies for a few selected
nuclei with those of Ref. \cite{Lalazissis99} obtained for $N_F = 12$.
We see that our results obtained with the $N_F = 12$ agree reasonably
well with those of Ref. \cite{Lalazissis99}. Though, the pairing
gaps used in our paper may be somewhat different.  Large differences
in the binding energy for the super-heavy nuclei in our work as
compared to that of Ref.  \cite{Lalazissis99} are mainly due to
the fact that we have used sufficiently large basis space (see also
Figs. \ref{fig:zpe0} and \ref{fig:zpe}).  All the results obtained within the
RMF models as presented below correspond to $N_F  = 18$  with $N_B = 20$.

\begin{figure*}
\resizebox{5.5in}{!}{ \includegraphics[]{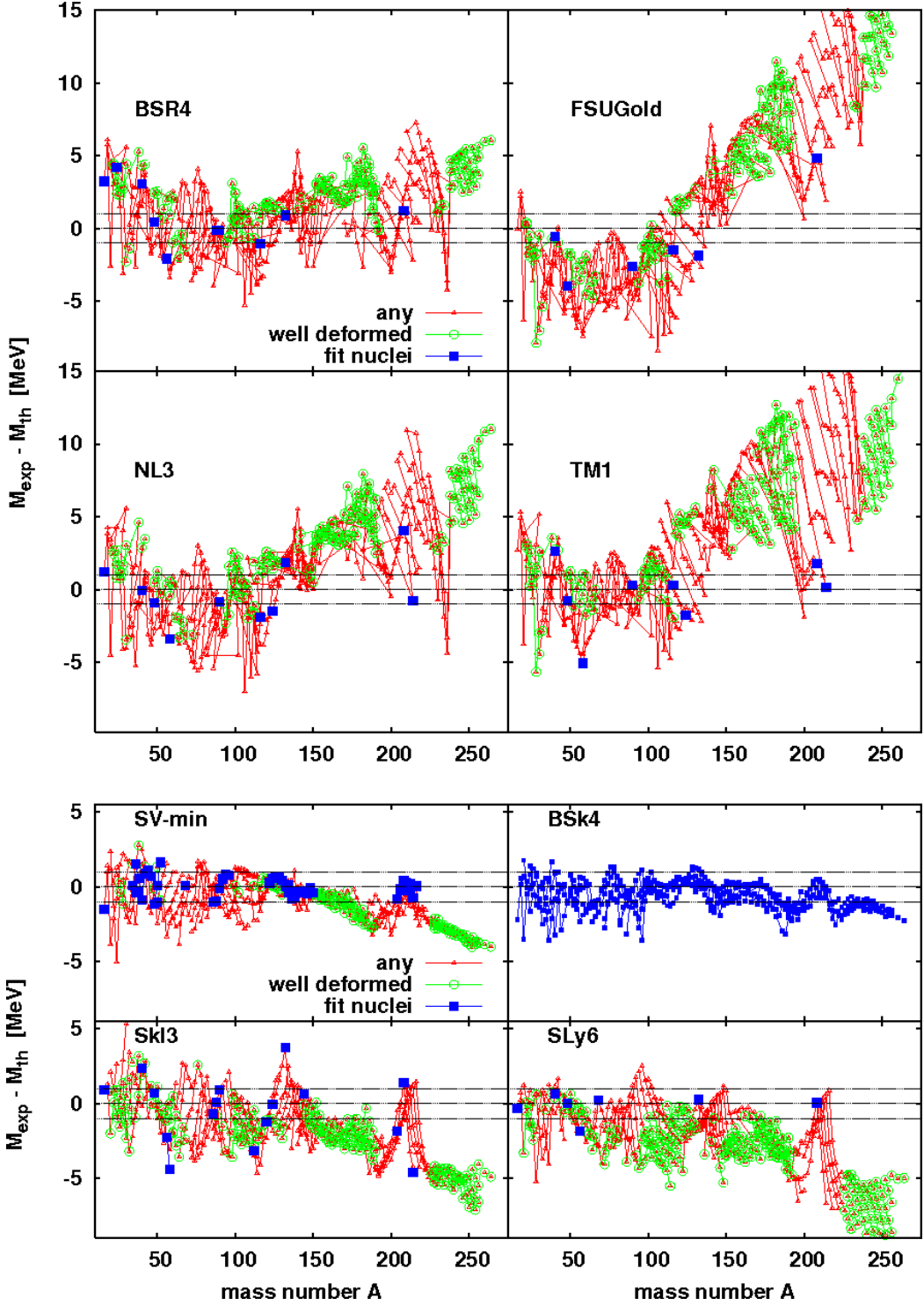}}
\caption{\label{fig:rmf_shf} (color online)
Errors in the binding energy 
versus the mass number
obtained for different parameterizations of the RMF (upper
block) and SHF (lower block). The
nuclei that were included in the fit are marked by filled squares,
well-deformed nuclei by open circles, and all others by triangles.
Binding energy error equal to  zero and $\pm 1$ MeV are indicated by faint
horizontal lines.
The corrections to the binding energies due to the pairing and quadrupole
correlations are included for all the cases (see text for detail).
 }
\end{figure*}

In Fig. \ref{fig:rmf_shf} we summarize the errors in binding energies (Eq.
\ref{eq:delM}) for all experimentally known even-even nuclei.  The upper
block collects the results from RMF and the lower block those from SHF.
As a first impression we note the rather small energy scale of a few
MeV for the error plots. All parameterizations shown here provide a good
description of known nuclei in the mass range $A=16$ -- 220.  There are,
of course, differences in detail to the extend that in each block the most
recent parameterizations perform visibly best. This concerns BSR4 in the
RMF block and BSk4 as well as SV-min in the SHF block. We have indicated
the present ``state of the art'' by $\pm 1$ MeV error bars and these
three most recent parameterizations stay close to this goal.  It is
also found as a general rule that robust spherical nuclei, denoted as
``fit nuclei'' in most parameterizations, are usually somewhat better
described than soft or deformed nuclei. At this point it is important to
emphasize that the still good description of soft and deformed nuclei is
achieved only with the rotational correction and one may reduce the error
even more by performing the full collective correlations \cite{Klu08a}.
We also calculate the  rms error for the binding energies using the
results for all the 513 even-even nuclei as considered in the present
work.  These rms errors as given in the parenthesis in the units of MeV
are: BSR4(2.6),FSUGold(6.5),NL3(3.8),TM1(5.9),SV-min(1.6),BSk4(1.1),
SkI3(2.6),SLy6(2.3).  Our rms error  for the case of BSk4 is higher than
$\sim 0.6$ MeV \cite{Gor03a}.  Because we use different rotational
correction, which probably has some effect for deformed nuclei.
Moreover, the basic paper for the BSk4 force uses a different recipe
for the cutoff of the pairing space. Their pairing band amounts to 15
MeV while we stay typically in the range of 5 MeV. That could have a
some effect on heavy nuclei.  Nevertheless, it may be emphasized that
these differences in detail do  not change the gross trends  for the
binding energy errors.  In particular trends for the binding energy
errors for the super-heavy nuclei presented in  Fig. \ref{fig:rmf_shf}
are very much similar to those of Ref. \cite{Gor03a}.

Having a closer look at the trends with mass number we see in all
cases a growing deviation from the zero line with increasing mass
number.  This trend shown here on a much larger data basis confirms
what had been observed in \cite{Klu09a}, namely that SHF drives to
underbinding of super-heavy nuclei while RMF shows just the opposite
trend. This trends become manifest in the region $A>220$, but develop
already for deformed nuclei in the region $150<A<220$. Within the SHF
parameterizations there are significant differences in the quantitative
level of the deviation. BSk4 manages best to minimize the unwanted
trend, achieved through the strong bias on binding energies, however,
other observables are compromised.  In the realm of the RMF the trend to
overbinding is even more dramatic. This is obvious in the traditional RMF,
parameterization NL3 in figure \ref{fig:rmf_shf}. Somewhat surprisingly,
the situation does not improve with $\omega$ self-couplings in TM1 and the
one more cross coupling in FSUGold. It is only BSR4 having the isoscalar
and isovector cross-couplings with the $\sigma$ fields which allows to
produce a much better overall trend in binding energies. These couplings
seem to be a crucial ingredient for success and it is to be noted that
SHF contains them in the isoscalar and isovector density-dependent terms.

\begin{figure*}
\resizebox{5.1in}{!}{ \includegraphics[]{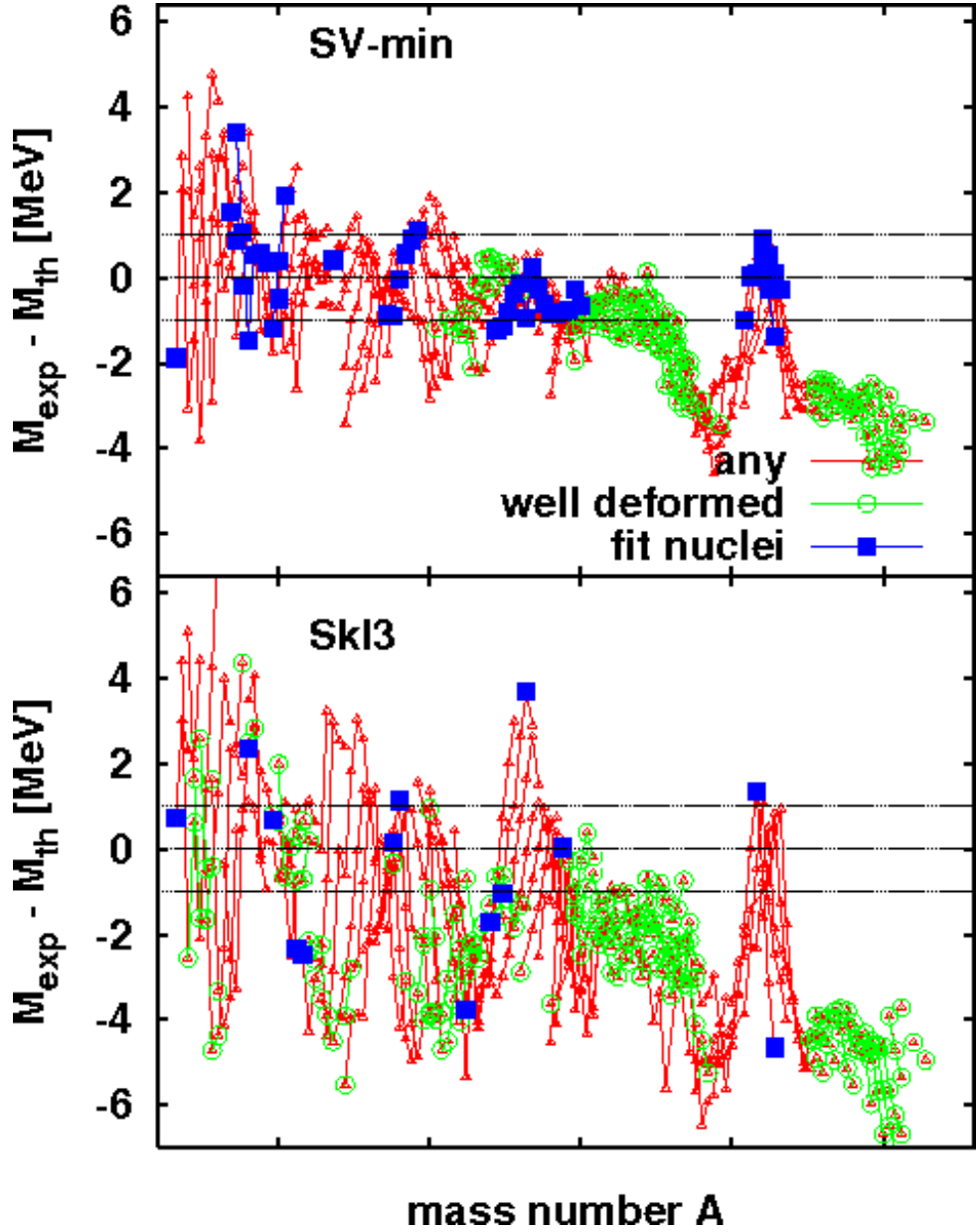}}
\caption{\label{fig:svmin-ski3} (color online)
Errors in the binding energies verses the mass number A.  The corrections
to the binding energies due to the pairing and quadrupole correlations are
evaluated approximately using Eqs. (\ref{eq:pair}) and (\ref{eq:erot}).
 }
\end{figure*}

We have looked into the possible consequences of the approximations used
in the present work  to evaluate the pairing and quadrupole correlation
energies for the RMF cases.  In Fig. \ref{fig:svmin-ski3}, we plot the SHF
results for the errors in the binding energies obtained for the SV-min and
SkI3 forces.  For both the cases, the pairing and quadrupole correlation
energies are calculated approximately using Eqs. (\ref{eq:pair}) and
(\ref{eq:erot}), respectively. Comparing these results with the
corresponding ones as shown in lower block of Fig. \ref{fig:rmf_shf},
which are obtained by treating the pairing and quadrupole correlations
microscopically, we find that the approximations used to evaluate the
pairing and quadrupole correlation energies do not affect the global
trends.  Finally, we compare our results for the BSR4 force for a few
heavy nuclei with those obtained using the density dependent meson
exchange, DDME2, force \cite{Lalazissis05}.  It appears that the BSR4
and DDME2 forces yield similar trends for the binding energy errors for
the heavy nuclei.

Comparing the three best performers, BSR4, BSk4 and SV-min, we see
that both models are approaching good control over the energies for
super-heavy elements. And yet, there remains an unresolved trend which
still is distinctively different between SHF and RMF. The reasons for
that are not yet clear. We try briefly to sort out possible
mechanisms. The defect seems to come from the deformation energy
because the large deviation develops with deformation. This, in turn,
localizes the differences in the modeling of the surface energy, and
most probably isovector surface energy (also called surface symmetry
energy) because heavier nuclei have naturally a larger asymmetry. A
direct relation between surface energies and model parameters has not
yet been established \cite{Rei06b}. At present we can only speculate
and try to figure out the terms which could have an influence.
Although one can map at lowest order $v/c$ the RMF into a Skyrme-like
functional including proper kinetic and spin-orbit terms
\cite{Sul07a}, there remain basic differences between the two classes
of models.  The spin-orbit and kinetic terms in the RMF carry
effectively a strong density dependence while the corresponding SHF
terms stay simply linear in $\rho$. This yields a different
performance with respect to spin orbit splitting and a different
density profile in super-heavy elements \cite{Rei02aR}, which both are
possible sources for different deformation energies.  Moreover, the
RMF produces a strong link between effective mass and spin-orbit force
where one of the consequences is that RMF models produce typically
very low effective masses in the range $m^*/m\approx 0.6$ while SHF
has much more freedom in the effective mass and in adjusting the
spin-orbit term independently.  It is certainly a worthwhile task for
future work to find out the mechanisms determining the trends as it
will shed light on the structure of both approaches, SHF and RMF, and
indicate missing pieces in the functionals.

\section{Summary}
In summary, we have calculated the binding energies, for all
experimentally known even-even nuclei, within the RMF and the SHF
models.  For the RMF models, we have used four different parameter
sets corresponding  to different functional forms for the non-linear
part of the Lagrangian density. The NL3 parameterization corresponds
to the Lagrangian density associated with the standard RMF model.
While the Lagrangian density for the  TM1, FSUGold and BSR4
parameterizations correspond to the extended RMF model  which include
contributions from the self-interaction of the $\omega$ meson and/or
the cross-interaction between $\sigma$, $\omega$ and $\rho$ mesons (see
Table \ref{tab:rmf}). For the SHF model we have used four parameter sets,
namely,  SkI3, SLy6, BSk4 and SV-min.  The form of the SHF functional
is basically the same for all four cases with the tiny difference that
SkI3 and SV-min include an isovector contribution to the spin-orbit
term \cite{Rei95a}.  We find that all parameterizations of the RMF and
SHF model considered provide a good description of known nuclei in the
mass range $A=16$ -- 220.  There are, of course, differences in detail to
the extent that the most recent parameterizations perform the best.
These parameterizations are BSR4 in the RMF model and BSk4 as well as
SV-min in the SHF model.

The binding energies for the super-heavy nuclei ($A > 220$) for the RMF
and SHF models are significantly different from the experimental data.
The  SHF drives to underbinding of super-heavy nuclei while RMF shows just
the opposite trend.  This trend on a much larger data basis confirms what
had been observed in \cite{Klu09a}.  The absolute errors in the binding
energies for the extended RMF model (BSR4) which includes contributions
from all the cross-interaction terms are comparable to that for the SHF
models. Other RMF models considered yield much larger values for the
absolute errors in the binding energies for the super-heavy nuclei.

\medskip

\noindent
Acknowledgment:
This work was supported by BMBF under contract no. 06~ER~142D.

\newpage
%\bibliography{1review}
%\end{document}
%\bibliography{systematics}

\begin{thebibliography}{33}
\expandafter\ifx\csname natexlab\endcsname\relax\def\natexlab#1{#1}\fi
\expandafter\ifx\csname bibnamefont\endcsname\relax
  \def\bibnamefont#1{#1}\fi
\expandafter\ifx\csname bibfnamefont\endcsname\relax
  \def\bibfnamefont#1{#1}\fi
\expandafter\ifx\csname citenamefont\endcsname\relax
  \def\citenamefont#1{#1}\fi
\expandafter\ifx\csname url\endcsname\relax
  \def\url#1{\texttt{#1}}\fi
\expandafter\ifx\csname urlprefix\endcsname\relax\def\urlprefix{URL }\fi
\providecommand{\bibinfo}[2]{#2}
\providecommand{\eprint}[2][]{\url{#2}}

\bibitem[{\citenamefont{Glendenning}(2000)}]{Gle00aB}
\bibinfo{author}{\bibfnamefont{N.~K.} \bibnamefont{Glendenning}},
  \emph{\bibinfo{title}{Compact Stars}} (\bibinfo{publisher}{Springer},
  \bibinfo{address}{New York}, \bibinfo{year}{2000}).

\bibitem[{\citenamefont{Stone and Reinhard}(2007)}]{Sto07aR}
\bibinfo{author}{\bibfnamefont{J.}~\bibnamefont{Stone}} \bibnamefont{and}
  \bibinfo{author}{\bibfnamefont{P.-G.} \bibnamefont{Reinhard}},
  \bibinfo{journal}{Prog. Part. Nucl. Phys.} \textbf{\bibinfo{volume}{58}},
  \bibinfo{pages}{587} (\bibinfo{year}{2007}).

\bibitem[{\citenamefont{Vautherin and Brink}(1972)}]{Vau72a}
\bibinfo{author}{\bibfnamefont{D.}~\bibnamefont{Vautherin}} \bibnamefont{and}
  \bibinfo{author}{\bibfnamefont{D.~M.} \bibnamefont{Brink}},
  \bibinfo{journal}{Phys. Rev. C} \textbf{\bibinfo{volume}{5}},
  \bibinfo{pages}{626} (\bibinfo{year}{1972}).

\bibitem[{\citenamefont{Bender et~al.}(2003)\citenamefont{Bender, Heenen, and
  Reinhard}}]{Ben03aR}
\bibinfo{author}{\bibfnamefont{M.}~\bibnamefont{Bender}},
  \bibinfo{author}{\bibfnamefont{P.-H.} \bibnamefont{Heenen}},
  \bibnamefont{and} \bibinfo{author}{\bibfnamefont{P.-G.}
  \bibnamefont{Reinhard}}, \bibinfo{journal}{Rev. Mod. Phys.}
  \textbf{\bibinfo{volume}{75}}, \bibinfo{pages}{121} (\bibinfo{year}{2003}).

\bibitem[{\citenamefont{Decharg{\'e} and Gogny}(1980)}]{Dec80a}
\bibinfo{author}{\bibfnamefont{J.}~\bibnamefont{Decharg{\'e}}}
  \bibnamefont{and} \bibinfo{author}{\bibfnamefont{D.}~\bibnamefont{Gogny}},
  \bibinfo{journal}{Phys. Rev. C} \textbf{\bibinfo{volume}{21}},
  \bibinfo{pages}{1568} (\bibinfo{year}{1980}).

\bibitem[{\citenamefont{Serot and Walecka}(1986)}]{Ser86aR}
\bibinfo{author}{\bibfnamefont{B.~D.} \bibnamefont{Serot}} \bibnamefont{and}
  \bibinfo{author}{\bibfnamefont{J.~D.} \bibnamefont{Walecka}},
  \bibinfo{journal}{Adv. Nucl. Phys.} \textbf{\bibinfo{volume}{16}},
  \bibinfo{pages}{1} (\bibinfo{year}{1986}).

\bibitem[{\citenamefont{Reinhard}(1989)}]{Rei89aR}
\bibinfo{author}{\bibfnamefont{P.-G.} \bibnamefont{Reinhard}},
  \bibinfo{journal}{Rep. Prog. Phys.} \textbf{\bibinfo{volume}{52}},
  \bibinfo{pages}{439} (\bibinfo{year}{1989}).

\bibitem[{\citenamefont{Ring}(1996)}]{Rin96aR}
\bibinfo{author}{\bibfnamefont{P.}~\bibnamefont{Ring}}, \bibinfo{journal}{Prog.
  Part. Nucl. Phys.} \textbf{\bibinfo{volume}{37}}, \bibinfo{pages}{193}
  (\bibinfo{year}{1996}).

\bibitem[{\citenamefont{Reinhard and Toepffer}(1994)}]{Rei94aR}
\bibinfo{author}{\bibfnamefont{P.-G.} \bibnamefont{Reinhard}} \bibnamefont{and}
  \bibinfo{author}{\bibfnamefont{C.}~\bibnamefont{Toepffer}},
  \bibinfo{journal}{Int. J. Mod. Phys. E} \textbf{\bibinfo{volume}{3}},
  \bibinfo{pages}{435} (\bibinfo{year}{1994}).

\bibitem[{\citenamefont{Navratil et~al.}(2000)\citenamefont{Navratil, Vary, and
  Barrett}}]{Nav00a}
\bibinfo{author}{\bibfnamefont{P.}~\bibnamefont{Navratil}},
  \bibinfo{author}{\bibfnamefont{J.~P.} \bibnamefont{Vary}}, \bibnamefont{and}
  \bibinfo{author}{\bibfnamefont{B.~R.} \bibnamefont{Barrett}},
  \bibinfo{journal}{Phys. Rev. C} \textbf{\bibinfo{volume}{62}},
  \bibinfo{pages}{054311} (\bibinfo{year}{2000}).

\bibitem[{\citenamefont{Wloch et~al.}(2005)\citenamefont{Wloch, Dean, Gour,
  Hjorth-Jensen, Kowalski, Papenbrok, and Piecuch}}]{Dea05a}
\bibinfo{author}{\bibfnamefont{M.}~\bibnamefont{Wloch}},
  \bibinfo{author}{\bibfnamefont{D.}~\bibnamefont{Dean}},
  \bibinfo{author}{\bibfnamefont{J.}~\bibnamefont{Gour}},
  \bibinfo{author}{\bibfnamefont{M.}~\bibnamefont{Hjorth-Jensen}},
  \bibinfo{author}{\bibfnamefont{K.}~\bibnamefont{Kowalski}},
  \bibinfo{author}{\bibfnamefont{T.}~\bibnamefont{Papenbrok}},
  \bibnamefont{and} \bibinfo{author}{\bibfnamefont{P.}~\bibnamefont{Piecuch}},
  \bibinfo{journal}{Phys. Rev. Lett.} \textbf{\bibinfo{volume}{94}},
  \bibinfo{pages}{212501} (\bibinfo{year}{2005}).

\bibitem[{\citenamefont{Navr\'atil et~al.}(2009)\citenamefont{Navr\'atil,
  Quaglioni, Stetcu, and Barrett}}]{Nav09aR}
\bibinfo{author}{\bibfnamefont{P.}~\bibnamefont{Navr\'atil}},
  \bibinfo{author}{\bibfnamefont{S.}~\bibnamefont{Quaglioni}},
  \bibinfo{author}{\bibfnamefont{I.}~\bibnamefont{Stetcu}}, \bibnamefont{and}
  \bibinfo{author}{\bibfnamefont{B.~R.} \bibnamefont{Barrett}},
  \bibinfo{journal}{J. Phys. G} \textbf{\bibinfo{volume}{36}},
  \bibinfo{pages}{083101} (\bibinfo{year}{2009}).

\bibitem[{\citenamefont{Kl\"upfel et~al.}(2009)\citenamefont{Kl\"upfel,
  Reinhard, B\"urvenich, and Maruhn}}]{Klu09a}
\bibinfo{author}{\bibfnamefont{P.}~\bibnamefont{Kl\"upfel}},
  \bibinfo{author}{\bibfnamefont{P.-G.} \bibnamefont{Reinhard}},
  \bibinfo{author}{\bibfnamefont{T.~J.} \bibnamefont{B\"urvenich}},
  \bibnamefont{and} \bibinfo{author}{\bibfnamefont{J.~A.}
  \bibnamefont{Maruhn}}, \bibinfo{journal}{Phys.Rev. C}
  \textbf{\bibinfo{volume}{79}}, \bibinfo{pages}{034310}
  (\bibinfo{year}{2009}).

\bibitem[{\citenamefont{Erler et~al.}(2010)\citenamefont{Erler, Kl\"upfel, and
  Reinhard}}]{Erl10a}
\bibinfo{author}{\bibfnamefont{J.}~\bibnamefont{Erler}},
  \bibinfo{author}{\bibfnamefont{P.}~\bibnamefont{Kl\"upfel}},
  \bibnamefont{and} \bibinfo{author}{\bibfnamefont{P.-G.}
  \bibnamefont{Reinhard}}, \bibinfo{journal}{J. Phys. G}
  \textbf{\bibinfo{volume}{37}}, \bibinfo{pages}{064001}
  (\bibinfo{year}{2010}).

\bibitem[{\citenamefont{Typel and Wolter}(1999)}]{Typ99a}
\bibinfo{author}{\bibfnamefont{S.}~\bibnamefont{Typel}} \bibnamefont{and}
  \bibinfo{author}{\bibfnamefont{H.~H.} \bibnamefont{Wolter}},
  \bibinfo{journal}{Nucl. Phys.} \textbf{\bibinfo{volume}{A656}},
  \bibinfo{pages}{331} (\bibinfo{year}{1999}).

\bibitem[{\citenamefont{Vretenar et~al.}(2005)\citenamefont{Vretenar,
  Lalazissis, Niksic, and Ring}}]{Vre05a}
\bibinfo{author}{\bibfnamefont{D.}~\bibnamefont{Vretenar}},
  \bibinfo{author}{\bibfnamefont{G.}~\bibnamefont{Lalazissis}},
  \bibinfo{author}{\bibfnamefont{T.}~\bibnamefont{Niksic}}, \bibnamefont{and}
  \bibinfo{author}{\bibfnamefont{P.}~\bibnamefont{Ring}},
  \bibinfo{journal}{Eur. Phys. J. A} \textbf{\bibinfo{volume}{25}},
  \bibinfo{pages}{555} (\bibinfo{year}{2005}).

\bibitem[{\citenamefont{Lalazissis et~al.}(1997)\citenamefont{Lalazissis,
  K{\"o}nig, and Ring}}]{Lal97a}
\bibinfo{author}{\bibfnamefont{G.~A.} \bibnamefont{Lalazissis}},
  \bibinfo{author}{\bibfnamefont{J.}~\bibnamefont{K{\"o}nig}},
  \bibnamefont{and} \bibinfo{author}{\bibfnamefont{P.}~\bibnamefont{Ring}},
  \bibinfo{journal}{Phys. Rev. C} \textbf{\bibinfo{volume}{55}},
  \bibinfo{pages}{540} (\bibinfo{year}{1997}).

\bibitem[{\citenamefont{Sugahara and Toki}(1994)}]{Sug94a}
\bibinfo{author}{\bibfnamefont{Y.}~\bibnamefont{Sugahara}} \bibnamefont{and}
  \bibinfo{author}{\bibfnamefont{H.}~\bibnamefont{Toki}},
  \bibinfo{journal}{Nucl. Phys. A} \textbf{\bibinfo{volume}{579}},
  \bibinfo{pages}{557} (\bibinfo{year}{1994}).

\bibitem[{\citenamefont{Todd-Rutel and Piekarewicz}(2005)}]{Tod05a}
\bibinfo{author}{\bibfnamefont{B.~G.} \bibnamefont{Todd-Rutel}}
  \bibnamefont{and}
  \bibinfo{author}{\bibfnamefont{J.}~\bibnamefont{Piekarewicz}},
  \bibinfo{journal}{Phys. Rev. Lett.} \textbf{\bibinfo{volume}{95}},
  \bibinfo{pages}{122501} (\bibinfo{year}{2005}).

\bibitem[{\citenamefont{Dhiman et~al.}(2007)\citenamefont{Dhiman, Kumar, and
  Agrawal}}]{Dhi07a}
\bibinfo{author}{\bibfnamefont{S.~K.} \bibnamefont{Dhiman}},
  \bibinfo{author}{\bibfnamefont{R.}~\bibnamefont{Kumar}}, \bibnamefont{and}
  \bibinfo{author}{\bibfnamefont{B.~K.} \bibnamefont{Agrawal}},
  \bibinfo{journal}{Phys. Rev. C} \textbf{\bibinfo{volume}{76}},
  \bibinfo{pages}{045801} (\bibinfo{year}{2007}).

\bibitem[{\citenamefont{Agrawal}(2010)}]{Agr10a}
\bibinfo{author}{\bibfnamefont{B.~K.} \bibnamefont{Agrawal}},
  \bibinfo{journal}{Phys. Rev. C} \textbf{\bibinfo{volume}{81}},
  \bibinfo{pages}{034323} (\bibinfo{year}{2010}).

\bibitem[{\citenamefont{M\"uller and Serot}(1996)}]{Mue96a}
\bibinfo{author}{\bibfnamefont{H.}~\bibnamefont{M\"uller}} \bibnamefont{and}
  \bibinfo{author}{\bibfnamefont{B.~D.} \bibnamefont{Serot}},
  \bibinfo{journal}{Nucl. Phys. A} \textbf{\bibinfo{volume}{606}},
  \bibinfo{pages}{508} (\bibinfo{year}{1996}).

\bibitem[{\citenamefont{Reinhard and Flocard}(1995)}]{Rei95a}
\bibinfo{author}{\bibfnamefont{P.-G.} \bibnamefont{Reinhard}} \bibnamefont{and}
  \bibinfo{author}{\bibfnamefont{H.}~\bibnamefont{Flocard}},
  \bibinfo{journal}{Nucl. Phys. A} \textbf{\bibinfo{volume}{584}},
  \bibinfo{pages}{467} (\bibinfo{year}{1995}).

\bibitem[{\citenamefont{Chabanat et~al.}(1998)\citenamefont{Chabanat, Bonche,
  Haensel, Meyer, and Schaeffer}}]{Cha98a}
\bibinfo{author}{\bibfnamefont{E.}~\bibnamefont{Chabanat}},
  \bibinfo{author}{\bibfnamefont{P.}~\bibnamefont{Bonche}},
  \bibinfo{author}{\bibfnamefont{P.}~\bibnamefont{Haensel}},
  \bibinfo{author}{\bibfnamefont{J.}~\bibnamefont{Meyer}}, \bibnamefont{and}
  \bibinfo{author}{\bibfnamefont{R.}~\bibnamefont{Schaeffer}},
  \bibinfo{journal}{Nucl. Phys. A} \textbf{\bibinfo{volume}{635}},
  \bibinfo{pages}{231} (\bibinfo{year}{1998}), \bibinfo{note}{{N}ucl. {P}hys.
  \textbf{A643}, 441(E)}.

\bibitem[{\citenamefont{Goriely et~al.}(2003)\citenamefont{Goriely, Samyn,
  Bender, and Pearson}}]{Gor03a}
\bibinfo{author}{\bibfnamefont{S.}~\bibnamefont{Goriely}},
  \bibinfo{author}{\bibfnamefont{M.}~\bibnamefont{Samyn}},
  \bibinfo{author}{\bibfnamefont{M.}~\bibnamefont{Bender}}, \bibnamefont{and}
  \bibinfo{author}{\bibfnamefont{J.~M.} \bibnamefont{Pearson}},
  \bibinfo{journal}{Phys. Rev. C} \textbf{\bibinfo{volume}{68}},
  \bibinfo{pages}{054325} (\bibinfo{year}{2003}).

\bibitem[{\citenamefont{Kl\"upfel et~al.}(2008)\citenamefont{Kl\"upfel, Erler,
  Reinhard, and Maruhn}}]{Klu08a}
\bibinfo{author}{\bibfnamefont{P.}~\bibnamefont{Kl\"upfel}},
  \bibinfo{author}{\bibfnamefont{J.}~\bibnamefont{Erler}},
  \bibinfo{author}{\bibfnamefont{P.-G.} \bibnamefont{Reinhard}},
  \bibnamefont{and} \bibinfo{author}{\bibfnamefont{J.~A.}
  \bibnamefont{Maruhn}}, \bibinfo{journal}{Eur. Phys. J. A}
  \textbf{\bibinfo{volume}{37}}, \bibinfo{pages}{343} (\bibinfo{year}{2008}).

\bibitem[{\citenamefont{Bender et~al.}(2000)\citenamefont{Bender, Rutz, Maruhn,
  and Reinhard}}]{Bender00}
\bibinfo{author}{\bibfnamefont{M.}~\bibnamefont{Bender}},
  \bibinfo{author}{\bibfnamefont{K.}~\bibnamefont{Rutz}},
  \bibinfo{author}{\bibfnamefont{J.~A.} \bibnamefont{Maruhn}},
  \bibnamefont{and} \bibinfo{author}{\bibfnamefont{P.-G.}
  \bibnamefont{Reinhard}}, \bibinfo{journal}{Eur. Phys. J.}
  \textbf{\bibinfo{volume}{A 7}}, \bibinfo{pages}{467} (\bibinfo{year}{2000}).


\bibitem[{\citenamefont{Bonche et~al.}(1985)\citenamefont{Bonche, Flocard,
  Heenen, Krieger, and Weiss}}]{Bon85a}
\bibinfo{author}{\bibfnamefont{P.}~\bibnamefont{Bonche}},
  \bibinfo{author}{\bibfnamefont{H.}~\bibnamefont{Flocard}},
  \bibinfo{author}{\bibfnamefont{P.-H.} \bibnamefont{Heenen}},
  \bibinfo{author}{\bibfnamefont{S.~J.} \bibnamefont{Krieger}},
  \bibnamefont{and} \bibinfo{author}{\bibfnamefont{M.~S.} \bibnamefont{Weiss}},
  \bibinfo{journal}{Nucl. Phys. A} \textbf{\bibinfo{volume}{443}},
  \bibinfo{pages}{39} (\bibinfo{year}{1985}).

\bibitem[{\citenamefont{Krieger et~al.}(1990)\citenamefont{Krieger, Bonche,
  Flocard, Quentin, and Weiss}}]{Kri90a}
\bibinfo{author}{\bibfnamefont{S.~J.} \bibnamefont{Krieger}},
  \bibinfo{author}{\bibfnamefont{P.}~\bibnamefont{Bonche}},
  \bibinfo{author}{\bibfnamefont{H.}~\bibnamefont{Flocard}},
  \bibinfo{author}{\bibfnamefont{P.}~\bibnamefont{Quentin}}, \bibnamefont{and}
  \bibinfo{author}{\bibfnamefont{M.~S.} \bibnamefont{Weiss}},
  \bibinfo{journal}{Nucl. Phys. A} \textbf{\bibinfo{volume}{517}},
  \bibinfo{pages}{275} (\bibinfo{year}{1990}).

\bibitem[{\citenamefont{Bender et~al.}(2000)\citenamefont{Bender, Rutz,
  Reinhard, and Maruhn}}]{Ben00c}
\bibinfo{author}{\bibfnamefont{M.}~\bibnamefont{Bender}},
  \bibinfo{author}{\bibfnamefont{K.}~\bibnamefont{Rutz}},
  \bibinfo{author}{\bibfnamefont{P.-G.} \bibnamefont{Reinhard}},
  \bibnamefont{and} \bibinfo{author}{\bibfnamefont{J.~A.}
  \bibnamefont{Maruhn}}, \bibinfo{journal}{Eur. Phys. J. A}
  \textbf{\bibinfo{volume}{8}}, \bibinfo{pages}{59} (\bibinfo{year}{2000}).


\bibitem[{\citenamefont{Ring and Schuck}(1980)}]{Rin80aB}
\bibinfo{author}{\bibfnamefont{P.}~\bibnamefont{Ring}} \bibnamefont{and}
  \bibinfo{author}{\bibfnamefont{P.}~\bibnamefont{Schuck}},
  \emph{\bibinfo{title}{The Nuclear Many-Body Problem}}
  (\bibinfo{publisher}{Springer}, \bibinfo{address}{New York},
  \bibinfo{year}{1980}).



\bibitem[{\citenamefont{Hagino et~al.}(2003)\citenamefont{Hagino, Bertsch, and
  Reinhard}}]{Hag03a}
\bibinfo{author}{\bibfnamefont{K.}~\bibnamefont{Hagino}},
  \bibinfo{author}{\bibfnamefont{G.}~\bibnamefont{Bertsch}}, \bibnamefont{and}
  \bibinfo{author}{\bibfnamefont{P.-G.} \bibnamefont{Reinhard}},
  \bibinfo{journal}{Phys. Rev. C} \textbf{\bibinfo{volume}{68}},
  \bibinfo{pages}{024306} (\bibinfo{year}{2003}).

\bibitem[{\citenamefont{Gambhir et~al.}(1990)\citenamefont{Gambhir, Ring, and
  Thimet}}]{Gambhir90}
\bibinfo{author}{\bibfnamefont{Y.~K.} \bibnamefont{Gambhir}},
  \bibinfo{author}{\bibfnamefont{P.}~\bibnamefont{Ring}}, \bibnamefont{and}
  \bibinfo{author}{\bibfnamefont{A.}~\bibnamefont{Thimet}},
  \bibinfo{journal}{Ann. Phys.(N.Y.)} \textbf{\bibinfo{volume}{198}},
  \bibinfo{pages}{132} (\bibinfo{year}{1990}).

\bibitem[{\citenamefont{Ring et~al.}(1997)\citenamefont{Ring, Gambhir, and
  Lalazissis}}]{Ring97}
\bibinfo{author}{\bibfnamefont{P.}~\bibnamefont{Ring}},
  \bibinfo{author}{\bibfnamefont{Y.~K.} \bibnamefont{Gambhir}},
  \bibnamefont{and} \bibinfo{author}{\bibfnamefont{G.~A.}
  \bibnamefont{Lalazissis}}, \bibinfo{journal}{Comp. Phys. Comm.}
  \textbf{\bibinfo{volume}{105}}, \bibinfo{pages}{77} (\bibinfo{year}{1997}).


\bibitem[{\citenamefont{Lalazissis et~al.}(1999)\citenamefont{Lalazissis,
  Raman, and Ring}}]{Lalazissis99}
\bibinfo{author}{\bibfnamefont{G.~A.} \bibnamefont{Lalazissis}},
  \bibinfo{author}{\bibfnamefont{S.}~\bibnamefont{Raman}}, \bibnamefont{and}
  \bibinfo{author}{\bibfnamefont{P.}~\bibnamefont{Ring}}, \bibinfo{journal}{At.
  Data Nucl. Data Tables} \textbf{\bibinfo{volume}{71}}, \bibinfo{pages}{1}
  (\bibinfo{year}{1999}).



\bibitem[{\citenamefont{Lalazissis et~al.}(2005)\citenamefont{Lalazissis,
  Niksic, Vretenar, , and Ring}}]{Lalazissis05}
\bibinfo{author}{\bibfnamefont{G.}~\bibnamefont{Lalazissis}},
  \bibinfo{author}{\bibfnamefont{T.}~\bibnamefont{Niksic}},
  \bibinfo{author}{\bibfnamefont{D.}~\bibnamefont{Vretenar}}, ,
  \bibnamefont{and} \bibinfo{author}{\bibfnamefont{P.}~\bibnamefont{Ring}},
  \bibinfo{journal}{Phys. Rev. C} \textbf{\bibinfo{volume}{71}},
  \bibinfo{pages}{024312} (\bibinfo{year}{2005}).



\bibitem[{\citenamefont{Reinhard et~al.}(2006)\citenamefont{Reinhard, Bender,
  Nazarewicz, and Vertse}}]{Rei06b}
\bibinfo{author}{\bibfnamefont{P.-G.} \bibnamefont{Reinhard}},
  \bibinfo{author}{\bibfnamefont{M.}~\bibnamefont{Bender}},
  \bibinfo{author}{\bibfnamefont{W.}~\bibnamefont{Nazarewicz}},
  \bibnamefont{and} \bibinfo{author}{\bibfnamefont{T.}~\bibnamefont{Vertse}},
  \bibinfo{journal}{Phys. Rev. C} \textbf{\bibinfo{volume}{73}},
  \bibinfo{pages}{014309} (\bibinfo{year}{2006}).

\bibitem[{\citenamefont{Sulaksono et~al.}(2007)\citenamefont{Sulaksono,
  Reinhard, B{\"u}rvenich, Hess, and Maruhn}}]{Sul07a}
\bibinfo{author}{\bibfnamefont{A.}~\bibnamefont{Sulaksono}},
  \bibinfo{author}{\bibfnamefont{P.~G.} \bibnamefont{Reinhard}},
  \bibinfo{author}{\bibfnamefont{T.~J.} \bibnamefont{B{\"u}rvenich}},
  \bibinfo{author}{\bibfnamefont{P.~O.} \bibnamefont{Hess}}, \bibnamefont{and}
  \bibinfo{author}{\bibfnamefont{J.~A.} \bibnamefont{Maruhn}},
  \bibinfo{journal}{Phys. Rev. Lett.} \textbf{\bibinfo{volume}{98}},
  \bibinfo{pages}{262501} (\bibinfo{year}{2007}).

\bibitem{Rei02aR}
{P.--G. Reinhard, M. Bender, and J. A. Maruhn},
{Comm. Mod. Phys. A } {\bf 2}, 177 (2002).


\end{thebibliography}

\newpage
\begin{table}
\caption{\label{tab:rmf}
Various  self-interaction and cross-interaction
terms present in the Lagrangian density associated with different
parameterizations of the  RMF
models. The index '1' and '0' is used to indicate whether or not the
corresponding term is included.
  }
%\begin{ruledtabular}
\begin{tabular}{ccccccc}
&
\multicolumn{3}{c}{\hspace*{-2em}Self-interaction\hspace*{-2em}}&
\multicolumn{3}{c}{Cross-interaction}\hspace*{-1em}\\
\cline{2-7}
 & $\sigma$& $\omega$& $\rho$& $\sigma\!-\!\omega$ 
&
%\hspace*{-0.2em}&\hspace*{-0.2em} 
$\sigma\!-\!\rho$
&
%\hspace*{-0.2em}&\hspace*{-0.2em} 
\!$\omega\!-\!\rho$\\
NL3& 1& 0& 0&0& 0& 0\\
TM1& 1& 1& 0&0& 0& 0\\
FSUGold& 1& 1& 0&0& 0& 1\\
BSR4& 1& 0& 0&1& 1& 1\\
\end{tabular}
%\end{ruledtabular}
\end{table}
%\newpage

\begin{center}
\begin{table}
\caption{\label {tab:comp}
Comparison of our results for the binding energies (in MeV)  for the
NL3 force with the corresponding ones given in Ref. \cite{Lalazissis99}.}
%\begin{ruledtabular}
\begin{tabular}{|cccc|}
\hline
\multicolumn{1}{|c}{}&
\multicolumn{1}{c}{Ref. \cite{Lalazissis99}}&
\multicolumn{2}{c|}{This work}\\
\cline{3-4}\\
\multicolumn{1}{|c}{Nucleus}&
\multicolumn{1}{c}{$N_F=12$}&
\multicolumn{1}{c}{$N_F=12$}&
\multicolumn{1}{c|}{$N_F=18$}\\
\hline
O16   &        &    128.82 &   128.84\\
Ca40 &  341.91 &   341.96  &    341.99\\
Ca48 & 415.07  &  415.11   &    415.06\\
Ni58 & 503.54  &  503.39   &     503.08\\
Zr90 & 783.41  &  783.08   &    783.06\\
Sn116& 986.44  &  986.15   &   986.79\\
Sn124& 1048.58 &   1048.15&   1048.49\\
Sn132& 1104.72 &   1104.81 &  1104.77\\
Pb208& 1640.16 &   1640.30 &   1640.50\\
Pb214& 1662.70 &   1662.02 &   1662.52\\
Th232& 1764.99 &   1764.12 &   1767.33\\
Cf248& 1857.17 &   1856.48 &   1861.51\\
Hs264&     &  1927.95 &   1934.68\\
\hline
\end{tabular}
%\end{ruledtabular}
\end{table}
\end{center}

\newpage

\begin{figure*}
\resizebox{6.1in}{!}{ \includegraphics[]{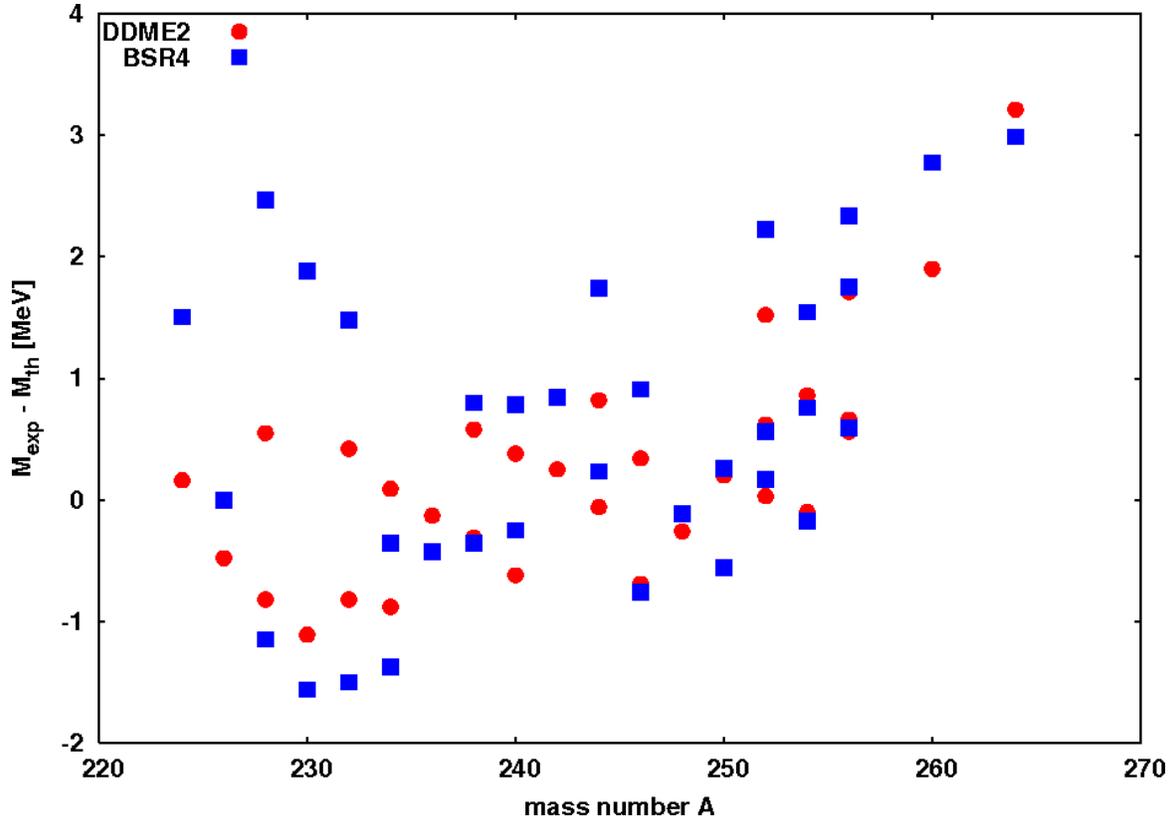}}
\caption{\label{fig:ddme2_bsr4} (color online)
Comparison of the bindign energy errors for the BSR4 force with those
for the DDME2 force for a few superheavy nuclei.
 }
\end{figure*}

\newpage

\end{document}